\documentclass[10pt,journal,compsoc]{IEEEtran}
\ifCLASSOPTIONcompsoc
  \usepackage[nocompress]{cite}
\else
\usepackage{cite}
\fi
\usepackage{geometry}
\usepackage{longtable}
\usepackage{tabularx}
\usepackage{lscape}
\usepackage{longtable}
\usepackage{graphicx}%
\usepackage{multirow}%
\usepackage{amsmath,amssymb,amsfonts}%
\usepackage{amsthm}%
\usepackage{mathrsfs}%
\usepackage{xcolor}%
\usepackage{textcomp}%
\usepackage{manyfoot}%
\usepackage{algorithm}%
\usepackage{algorithmicx}%
\usepackage{algpseudocode}%
\usepackage{listings}%
\usepackage{multicol}
\usepackage{lscape}
\usepackage{array} 
\usepackage{hyperref}
\usepackage{amsfonts}
\usepackage{amsmath} 
\usepackage{pdflscape}
\usepackage{longtable}
\usepackage{graphicx}
\usepackage{algorithm}
\usepackage{algpseudocode}
\usepackage{booktabs}
\usepackage{multirow} 
\usepackage{hyperref}
\usepackage{url}
\usepackage{multirow}
\usepackage{amsmath}
\usepackage{pgfplots}
\pgfplotsset{compat=1.17}
\usepackage{listings}
\usepackage{xcolor}
\usepackage{orcidlink}
\ifCLASSINFOpdf

\else

\fi

\begin{document}

\title{Large Language Models for EEG: A Comprehensive Survey and Taxonomy}

\author{
    Naseem Babu\,\orcidlink{0000-0001-7145-3531}\thanks{Department of Computer Science and Engineering, Indian Institute of Technology Patna, Bihta, Patna, India, 801106. Email: naseem\_2021cs22@iitp.ac.in}
    \and 
    Jimson Mathew\,\orcidlink{0000-0001-8247-9040}\thanks{Department of Computer Science and Engineering, Indian Institute of Technology Patna, Bihta, Patna, India, 801106. Email: jimson@iitp.ac.in}
    \and 
    and A. P. Vinod\,\orcidlink{0000-0001-9408-1275}\thanks{Infocomm Technology Cluster, Singapore Institute of Technology,
10 Dover Drive, Singapore 138683 Email: vinod.prasad@singaporetech.edu.sg}
}



\IEEEtitleabstractindextext{%
\begin{abstract}
The growing convergence between Large Language Models (LLMs) and electroencephalography (EEG) research is enabling new directions in neural decoding, brain-computer interfaces (BCIs), and affective computing. This survey offers a systematic review and structured taxonomy of recent advancements that utilize LLMs for EEG-based analysis and applications. We organize the recent studies into four categories: (1) LLM-inspired foundation models for EEG analysis, (2) EEG-to-language decoding, (3) cross-modal generation including image and 3D object synthesis, and (4) clinical applications and dataset management tools. The survey highlights how transformer-based architectures adapted through fine-tuning, few-shot, and zero-shot learning have enabled EEG-based models to perform complex tasks such as natural language generation, semantic interpretation, and diagnostic assistance. By presenting a structured overview of the employed models and application domains, this survey establishes a comprehensive framework to advance neural signal analysis through the application of language models.
\end{abstract}

}

\maketitle
\IEEEdisplaynontitleabstractindextext
\IEEEpeerreviewmaketitle
\begin{table*}[ht!]

\caption{Acronyms and abbreviations used in this survey}
\centering
\begin{tabular}{lccl}
\toprule
\hline
\textbf{Acronym} & & & \textbf{Full Form} \\
\midrule
LLM & & &  Large Language Model \\
EEG & & &  Electroencephalography \\
BCI & & &  Brain-Computer Interface \\
MI & & &  Motor Imagery\\
BERT & & &  Bidirectional Encoder Representations from Transformers \\
BART & & &  Bidirectional and Auto-Regressive Transformers \\
GPT & & &  Generative Pre-trained Transformer \\
BELT & & &  Bridging Electroencephalogram with Language Transformers \\
CET-MAE & & &  Contrastive EEG-Text Masked Autoencoder \\
LLaMA & & &  Large Language Model Meta AI \\
FLAN-T5 & & &  Fine-tuned Language Net T5 \\
LaBraM & & &  Large Brain Model \\
EEGPT & & &  Electroencephalography Pretrained Transformer \\
LCM & & &  Large Cognition Model \\
EEGFormer & & &  EEG Transformer \\
Neuro-GPT & & &  Neurological Generative Pre-trained Transformer \\
EEG2TEXT & & &  Electroencephalogram to Text \\
SEE & & &  Semantically Aligned EEG-to-Text Translation \\
EEGTrans & & &  Electroencephalogram Transformer \\
ELM & & &  EEG-Language Model \\
AdaCT & & &  Adapter for Converting Time Series \\
E2T-PTR & & &  EEG-to-Text using Pretrained Transferable Representations \\
EEG-CLIP & & &  EEG-based Contrastive Language-Image Pretraining \\
HMLLM & & &  Hypergraph Multi-modal Large Language Model \\
MLM & & &  Masked Language Modeling \\
ALM & & &  Autoregressive Language Modeling \\
NLP & & &  Natural Language Processing \\
FCM & & &  Fuzzy C-Means \\
FJM & & &  Fuzzy J-Means \\
\hline
\bottomrule
\end{tabular}
\label{tab:acronyms}
\end{table*}

\section{Introduction}
\begin{figure}[ht!]
\centering
\includegraphics[width=7cm]{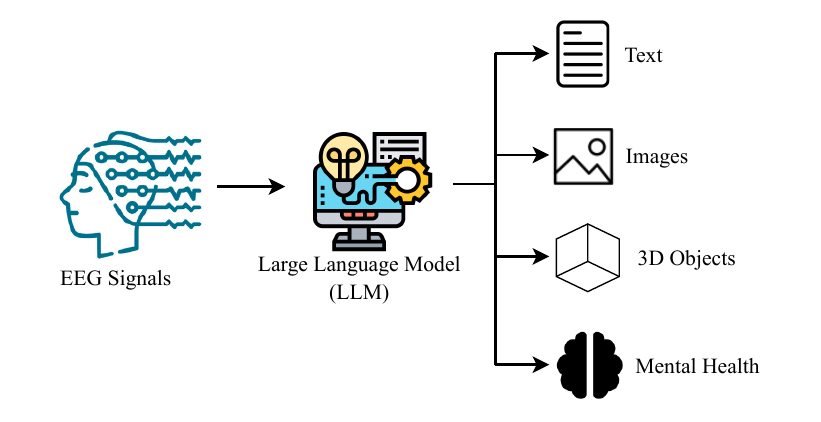}
\caption{EEG signals are processed by a large language model that interprets the underlying neural patterns. Depending on the task, the system generates outputs such as text, images, 3D reconstructions, or diagnostic insights.}
\label{overview}
\end{figure}

The growing use of language models is shaping new developments in both neuroscience and artificial intelligence. Language models such as GPT, BERT \cite{bert04, bert03, bert02}, and their multimodal variants have shown remarkable success across tasks involving natural language processing and understanding, generation, and even vision-language fusion. The transformer-based architecture enables these models to effectively process sequential data, contributing to their success across a range of domains beyond natural language processing. Alongside these developments, the field of electroencephalogram (EEG) \cite{bci01, bci02} based brain signal processing has seen substantial growth, driven by its promising applications in brain-computer interfaces (BCIs) \cite{eeg01, eeg03, eeg04}, cognitive assessment, mental health monitoring, and neurological diagnostics. The convergence of large language models and EEG signals offers a unique opportunity to bridge artificial intelligence and neuroscience. Leveraging LLMs enables the translation of brain activity into meaningful outputs such as text, images, 3D objects, and diagnostic insights \cite{entry29, entry16, entry25, entry14}, as illustrated in Fig.~\ref{overview}. The acronyms used throughout this paper are summarized in TABLE~\ref{tab:acronyms}.

\subsection{EEG and Language Models}
Despite recent advancements, EEG signal processing remains highly challenging due to the noisy nature of non-invasive recordings, which are often contaminated by artifacts such as eye movements, muscle activity, and external interference \cite{artifact01, artifact02, artifact03}. Additionally, these signals show inter-subject and intra-subject variability, complicating generalization across individuals and recording sessions \cite{subject_variability01, subject_variability02}. Lack of datasets poses a challenge for effective model training and evaluation. As a result, many existing machine and deep learning approaches remain confined to single-task, single-dataset settings, limiting their scalability and robustness. To address these challenges, more flexible and scalable learning frameworks are needed, and language models \cite{llm01}, with their ability to capture complex dependencies and adapt across modalities, offer promising capabilities to overcome these limitations.

\begin{itemize}
\item \textbf{Self-attention for spatiotemporal modeling:}
The self-attention mechanism in language models enables dynamic weighting of input features and effectively captures long-range dependencies, which are essential for modeling the spatiotemporal dynamics of signals \cite{attention, llm03, llm05}.   

\item \textbf{Transformer adaptability to neural signals:}
The flexibility of the transformer architecture allows it to effectively handle multichannel, multi-timescale inputs by learning embedding-based representations that capture both spatial and temporal characteristics of signals \cite{boka04, boka05}.

\item \textbf{Cross-modal generation capabilities:}
Combining EEG signals with multimodal language models enables novel applications such as EEG-to-text generation, brain-driven image synthesis, and exploratory 3D reconstruction. These advances expand the possibilities for interpreting and visualizing brain activity \cite{entry27, entry16, entry25}.
\end{itemize}


\subsection{Scope and Contributions}
This survey reviews recent advancements combining EEG signal processing and large language models (LLMs), focusing on applications such as EEG-to-text decoding, emotion recognition, multimodal generation, and mental health diagnostics.

The main contributions of this work are as follows:

\begin{enumerate}
\item Introduces a structured taxonomy covering four domains: foundation models, decoding, cross-modal generation, and clinical applications.

\item Summarizes recent studies through comprehensive tables detailing the tasks, datasets, methods, and model types used.

\item Highlights how different LLM architectures have been adapted for EEG-related tasks and outlines their roles across various applications.

\item Identifies key challenges in this emerging field, including data scarcity, interpretability, and real-time deployment, and outlines directions for future work.
\end{enumerate}

The remainder of this paper is organized as follows: Section~\ref{background} provides background on electroencephalography (EEG) signals and large language models (LLMs); Section~\ref{taxonomy_section} presents a taxonomy of language models applications in EEG analysis; Section~\ref{adaptation} outlines model adaptation strategies in EEG analysis; Section~\ref{studies} gives a detailed discussion on recent studies; Section~\ref{future} discusses future research directions; and Section~\ref{conclusion} concludes the paper.

\section{Background}\label{background}
\subsection{Electroencephalography (EEG) Signals}
EEG is a non-invasive neuroimaging technique that records electrical activity in the brain with electrodes placed on the scalp \cite{eeg01, eeg03, eeg04}. These signals reflect the collective electrical activity of cortical neurons and are captured as voltage fluctuations over time. These signals are inherently multichannel time series, typically sampled at high frequencies (e.g., 128-1024 Hz) across arrays of electrodes arranged according to standardized systems such as the international 10-20 or 10-10 montage. Each electrode provides spatially localized insight into cortical function, supporting both region-specific and network-level analyses. These signals are often decomposed into frequency bands: delta (0.5-4 Hz), theta (4-8 Hz), alpha (8-13 Hz), beta (13-30 Hz), and gamma (30-100 Hz) \cite{eeg03, eeg04}, where each band is associated with different cognitive or physiological states. For example, alpha waves are linked to relaxed wakefulness, while beta waves correspond to focused attention and active mental engagement. Despite their versatility, these signals are highly susceptible to various artifacts, such as those arising from eye blinks, facial muscle movements, and environmental interference, making preprocessing and artifact rejection essential before further analysis \cite{artifact01, artifact02, artifact03}. EEG is widely used across various domains, including Brain-Computer Interfaces (BCIs) \cite{bci01, bci02, bci04}, where it enables direct communication between the brain and external devices using mental states such as motor imagery \cite{motor_imagery01, motor_imagery02} or steady-state visual evoked potentials (SSVEPs) \cite{SSVEP01, SSVEP02}; clinical diagnosis of neurological conditions like epilepsy, sleep disorders, and encephalopathies; and applications in cognitive monitoring, emotion recognition, neurofeedback, and neuroergonomics, supporting real-time assessment of mental workload, attention, and affective states in both laboratory and real-world settings.

\subsection{Large Language Models (LLMs)}
LLMs are deep neural architectures designed to model and generate human-like language \cite{llm01,llm05,llm06}. The backbone of language models is the transformer architecture, introduced by Vaswani et al. in 2017 \cite{llm07}. The key innovation in transformers is the self-attention mechanism, which allows the model to assign dynamic importance to different positions in a sequence, thereby capturing both local and global dependencies. Unlike recurrent neural networks, transformers are highly parallelizable and scalable, enabling them to process longer sequences efficiently. Two major pretraining paradigms have emerged in large language models:

\begin{enumerate}
\item \textbf{Masked Language Modeling (MLM):}
This approach involves masking a portion of the input tokens and training the model to predict the masked content using both left and right context. BERT \cite{han2021pre, bert01} is a well-known example of MLM, leveraging bidirectional modeling to learn rich contextual embeddings that are particularly effective for classification and semantic understanding tasks.

\item \textbf{Autoregressive Language Modeling (ALM):}
This approach trains a model to predict the next token in a left-to-right sequence, enabling the generation of coherent long-form text. GPT \cite{han2021pre} is a prominent example of this method, which is well-suited for applications such as dialogue systems, document completion, and EEG-to-text decoding.
\end{enumerate}

With the continued progress of large language models, there has been a shift toward supporting diverse input modalities and learning objectives. These models have evolved from text-only systems to multimodal frameworks (e.g., Flamingo, LLaVA, GPT-4V) \cite{alayrac2022flamingo, liu2023llava} capable of processing combinations of text, images, and audio. Instruction-tuned variants such as FLAN-T5 and ChatGPT demonstrate how prompting can guide them to perform a wide range of tasks with little or no task-specific fine-tuning \cite{wei2021finetuned}. This is particularly promising for brain signal analysis, where labeled data is often limited and zero or few-shot prompting can leverage domain-specific cues. These advancements have extended the role of LLMs beyond traditional language tasks, enabling applications in neural decoding, cognitive state modeling, and multimodal brain-machine interfaces.

\section{A Taxonomy of LLM Applications in EEG Analysis}\label{taxonomy_section}
To provide a clear and structured overview of how large language models are being used in EEG analysis, we present a taxonomy that categorizes recent studies based on their primary objective, methodology, and output modality. The surveyed works are organized into four categories:

\begin{enumerate}
\item \textbf{LLM-Inspired EEG Foundation Models:}
This category includes approaches that adopt transformer-based architectures to learn general-purpose EEG representations that can be transferred across multiple tasks. Inspired by foundation models in natural language processing, these models aim to capture spatiotemporal features through large-scale pretraining. Two primary modeling strategies are commonly used: masked modeling and autoregressive modeling.

\item \textbf{EEG-to-Language Decoding:}
This category focuses on generating natural language from brain signals. Techniques include decoder-based approaches that use neural embeddings as prompts or inputs to text decoders, semantic alignment methods that map signal representations to language embeddings in a shared space, and instruction-tuned models that leverage prompt-based learning for interpretation with minimal supervision.

\item \textbf{Cross-Modal EEG Generation:}
This category explores the translation of brain activity into other modalities such as images, text, or 3D objects, enabling richer interpretations of neural signals and supporting applications like brain-to-image synthesis and brain-to-text decoding.

\item \textbf{Clinical Applications and Dataset Tools:}
This category encompasses applications such as emotion recognition, mental health diagnostics, and motor imagery classification. It also includes efforts focused on dataset tools and report clustering for automated annotation, as well as cognitive and reading analysis that links brain signals to attention and comprehension levels.
\end{enumerate}

\begin{figure*}[ht!]
    \centering
    \includegraphics[width=15cm]{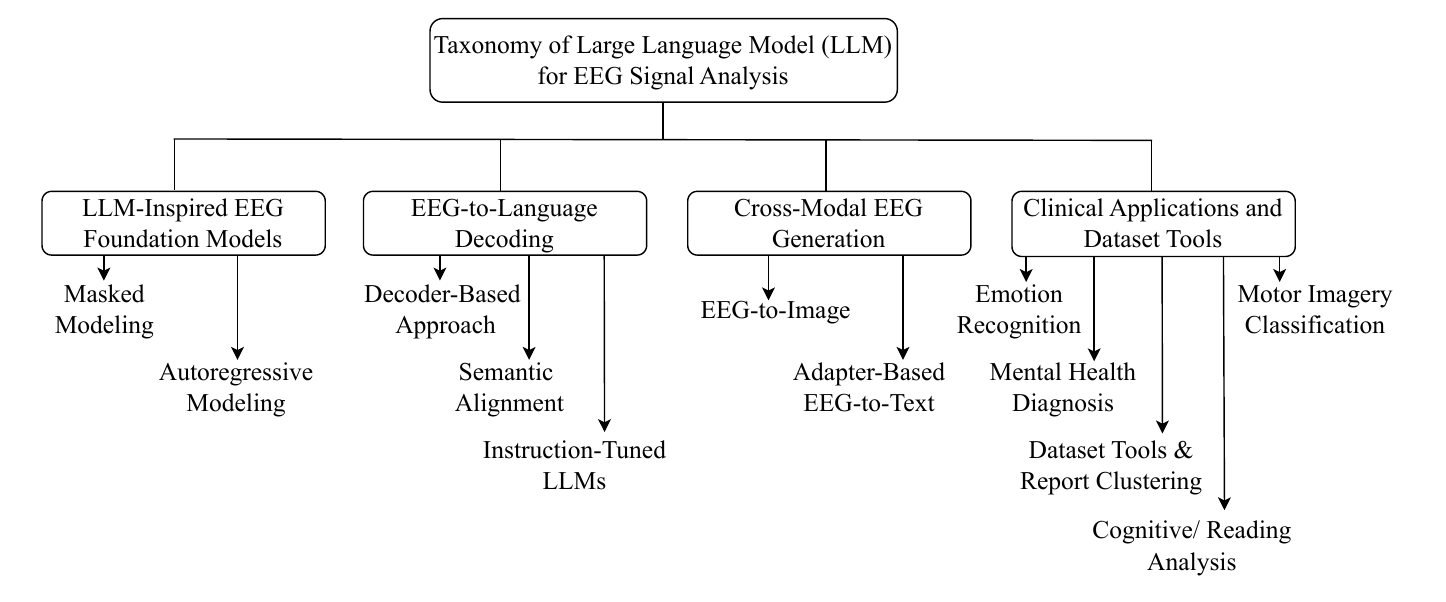}
    \caption{A taxonomy of large language model applications in EEG signal analysis, organized into four categories: foundation models, language decoding, cross-modal generation, and clinical applications.}
    \label{taxonomy}
\end{figure*}

\begin{figure}[ht!]
    \centering
    \includegraphics[width=7cm]{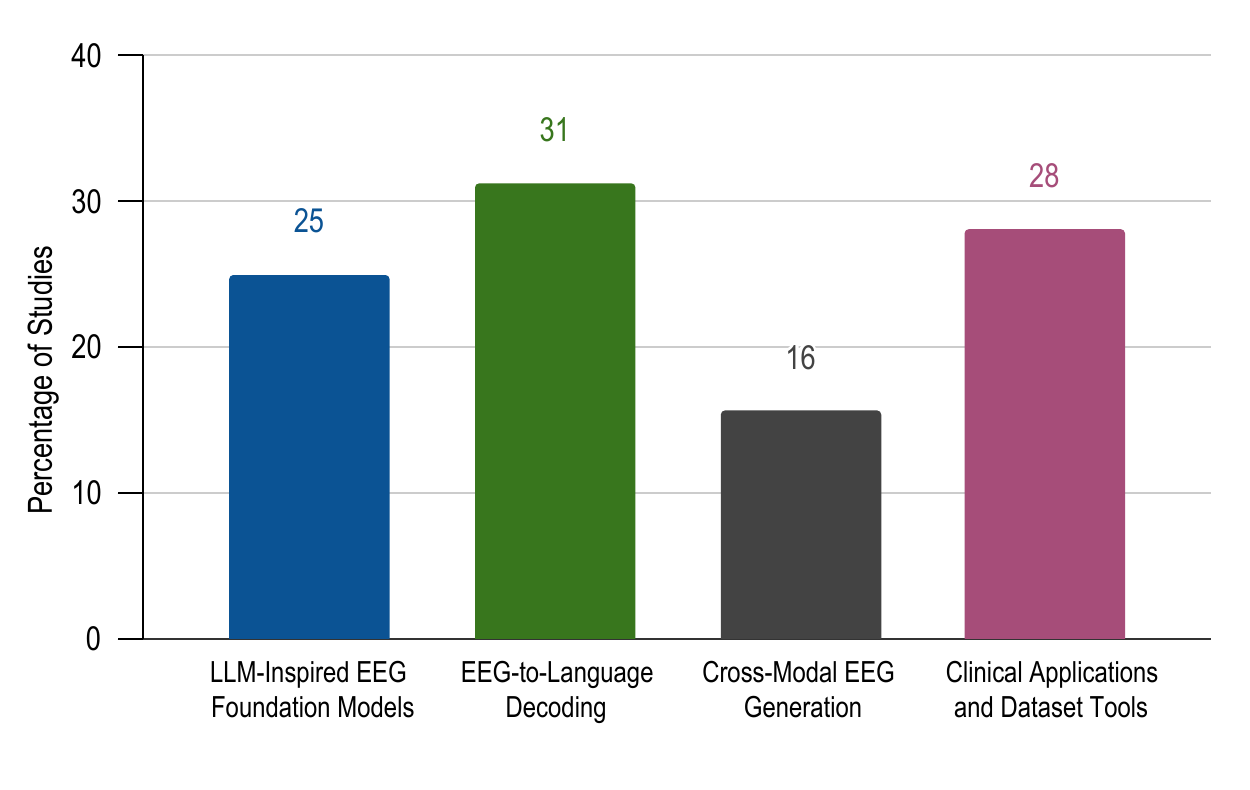}
        \caption{Distribution of EEG-LLM studies across four domains, reflecting current research trends and areas of emphasis in the application of language models to EEG signal analysis.}
    \label{review_distribution}
\end{figure}

Fig.~\ref{taxonomy} provides a visual overview of this classification, highlighting the functional contributions of language models in EEG processing and their diverse applications across related tasks. To better understand how language models are being applied to EEG analysis, Fig.~\ref{review_distribution} illustrates the distribution of existing studies, emphasizing the current focus areas within this emerging field.

\section{LLM Adaptation Strategies in EEG Analysis}\label{adaptation}
As LLMs find growing use in EEG analysis, various strategies have emerged to adapt them for related tasks. Based on recent studies, these strategies can be categorized into three types: fine-tuning, few-shot learning, and zero-shot learning.

\subsection{Fine-Tuning}
\begin{figure}[ht!]
\centering
\includegraphics[width=7cm]{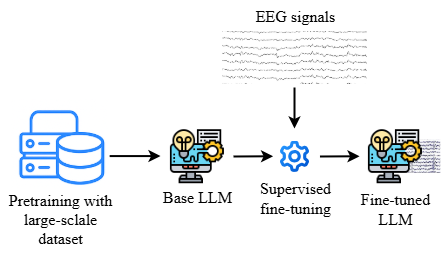}
\caption{Workflow illustrating the fine-tuning of a large language model for EEG analysis. A pretrained LLM is adapted using labeled data to perform domain-specific tasks.}
\label{finetune}
\end{figure}

Fine-tuning is a widely used method for adapting language models to domain-specific tasks. In the context of neural signal analysis, this involves retraining a pretrained LLM using labeled EEG data, enabling the model to learn domain-relevant patterns and perform specialized downstream tasks such as classification, interpretation, or generation \cite{fine_tuning}. Depending on the available resources and task requirements, fine-tuning can range from updating the entire model to using lightweight methods like prefix tuning. In some cases, only small additional modules are trained while keeping the core LLM unchanged. Despite requiring labeled data and moderate computational effort, fine-tuning remains a powerful and flexible approach for EEG applications. In Thought2Text \cite{entry29} uses instruction-tuned language models such as LLaMA, Mistral, and Qwen2.5, which are first fine-tuned on multimodal datasets (e.g., image-text pairs) and then applied to EEG embeddings for generating open-ended textual descriptions. AdaCT \cite{entry23} introduces a plug-and-play approach using adapters that transform EEG signals into pseudo-text or pseudo-image formats, enabling fine-tuning with pretrained vision or language models. Similarly, BELT-2 \cite{entry02} adopts prefix tuning, a parameter-efficient strategy where learnable prefix vectors are prepended to LLM layers to align EEG representations with GPT-style decoders for EEG-to-text translation. An illustrative overview of the fine-tuning process, including its integration with EEG inputs, is presented in Fig.~\ref{finetune}.

\subsection{Zero-Shot Learning}
\begin{figure}[ht!]
\centering
\includegraphics[width=7cm]{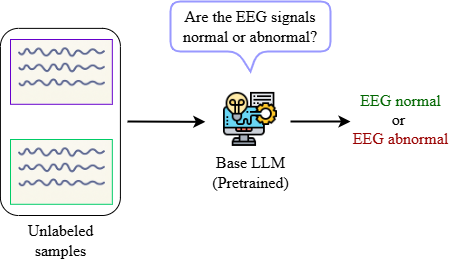}
\caption{Illustration of a zero-shot approach using a pretrained language model, where unlabeled EEG signals are provided along with a task-specific prompt (e.g., “Are the EEG signals normal or abnormal?”). The model performs inference without task-specific training or examples, directly predicting the outcome based on the prompt.}
\label{zeroshot}
\end{figure}

Zero-shot learning pushes generalization to the extreme by allowing a pretrained language model to perform tasks without being exposed to any task-specific training examples. Instead, the model is prompted using only natural language instructions or semantic cues \cite{zeroshot01, zeroshot02}. In the context of EEG analysis, this approach enables the transfer of knowledge to new tasks, patients, or datasets without the need for additional labeled data or retraining. Several EEG-based applications have explored this paradigm, like EEG-CLIP \cite{entry07}, which aligns time-series data with clinical text reports through contrastive learning, enabling both zero-shot classification and retrieval via a shared embedding space. Similarly, the Video-SME framework \cite{entry18} incorporates a Hypergraph Multimodal Large Language Model (HMLLM) to jointly interpret EEG and eye-tracking data during video viewing, using zero-shot inference to uncover semantic associations between modalities. Additionally, components of the Mental Health Classifier and Clinical Report Clustering systems \cite{entry14} apply zero-shot language model reasoning to analyze patient states or group clinical documents, without requiring retraining on those specific target tasks. An overview of the zero-shot learning workflow is illustrated in Fig.~\ref{zeroshot}.

\subsection{Few-Shot Learning}
\begin{figure}[ht!]
\centering
\includegraphics[width=7cm]{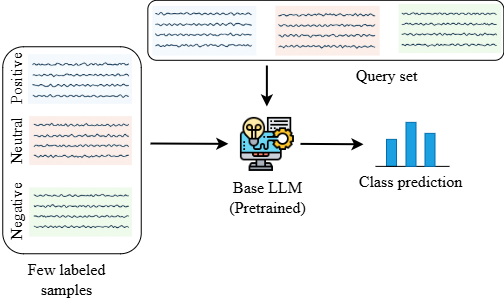}
\caption{Illustration of few-shot learning for emotion classification using a pretrained large language model. A small support set with labeled samples from three emotion classes, positive, neutral, and negative, is included in-context prompt. The model receives unlabeled query samples and infers their emotion class by reasoning over the support set, producing a probability distribution across the three categories.}
\label{fewshort}
\end{figure}

Few-shot learning leverages the ability of language models to generalize from a small number of labeled examples embedded in the prompt at inference time, through a process known as in-context learning. Unlike fine-tuning, few-shot learning does not involve updating the model's parameters, making it computationally efficient and particularly suitable for low-resource environments \cite{fewshot01, fewshot02}. In EEG analysis, few-shot learning has been applied to various tasks involving multimodal data. The Multimodal Mental Health Classifier \cite{entry14} employs few-shot prompting to perform emotion and depression classification using EEG signals alongside facial and textual features. In this setup, models like GPT-4 are prompted with carefully structured example pairs, allowing them to generalize to new, unseen inputs. Similarly, EEG Emotion Copilot \cite{entry06} applies few-shot style querying during inference to generate adaptive feedback and emotion summaries based on input. Overall, few-shot approaches are highly effective in scenarios with limited labeled data or when rapid prototyping is needed without the overhead of full model retraining. A schematic representation of the few-shot learning is provided in Fig.~\ref{fewshort}.

\section{EEG LLM Based Studies}\label{studies}
\begin{figure*}[ht!]
    \centering
      \includegraphics[width=15cm]{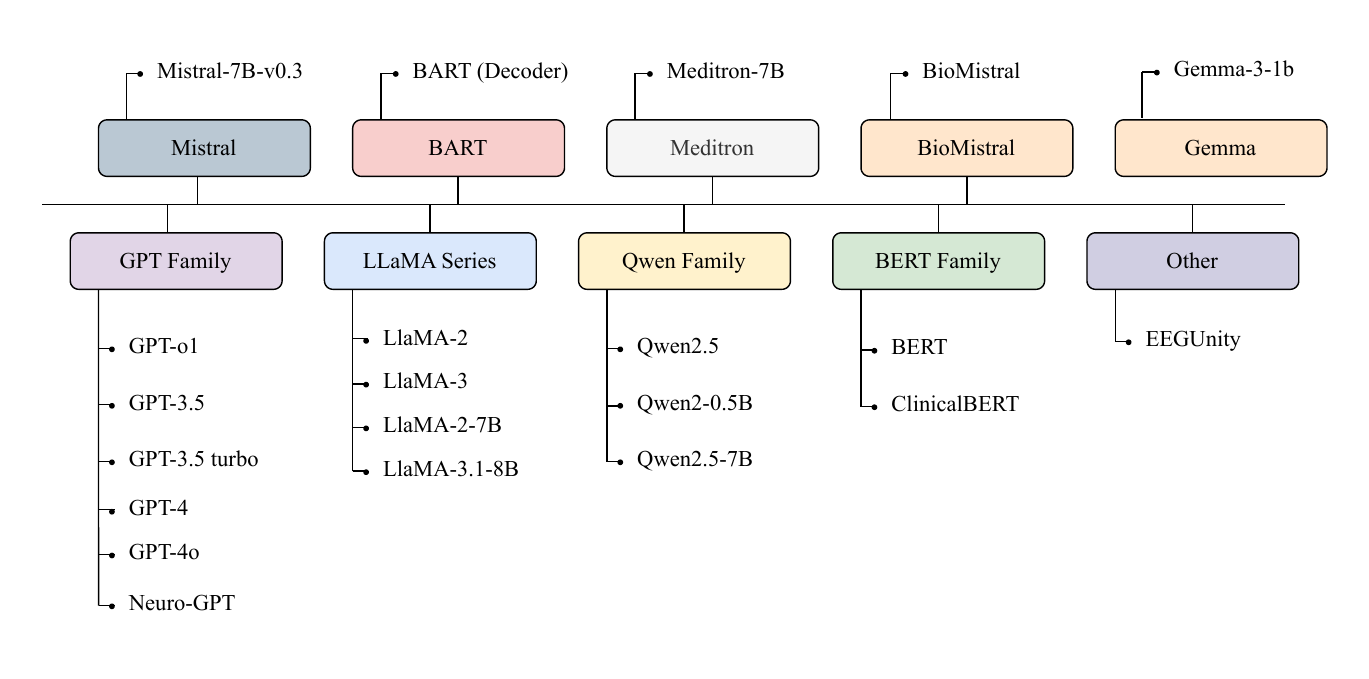}
    \caption{Summary of large language models (LLMs) used in EEG studies.}

    \label{llm_family}
\end{figure*}

\begin{figure*}[ht!]
    \centering
    \includegraphics[width=15cm]{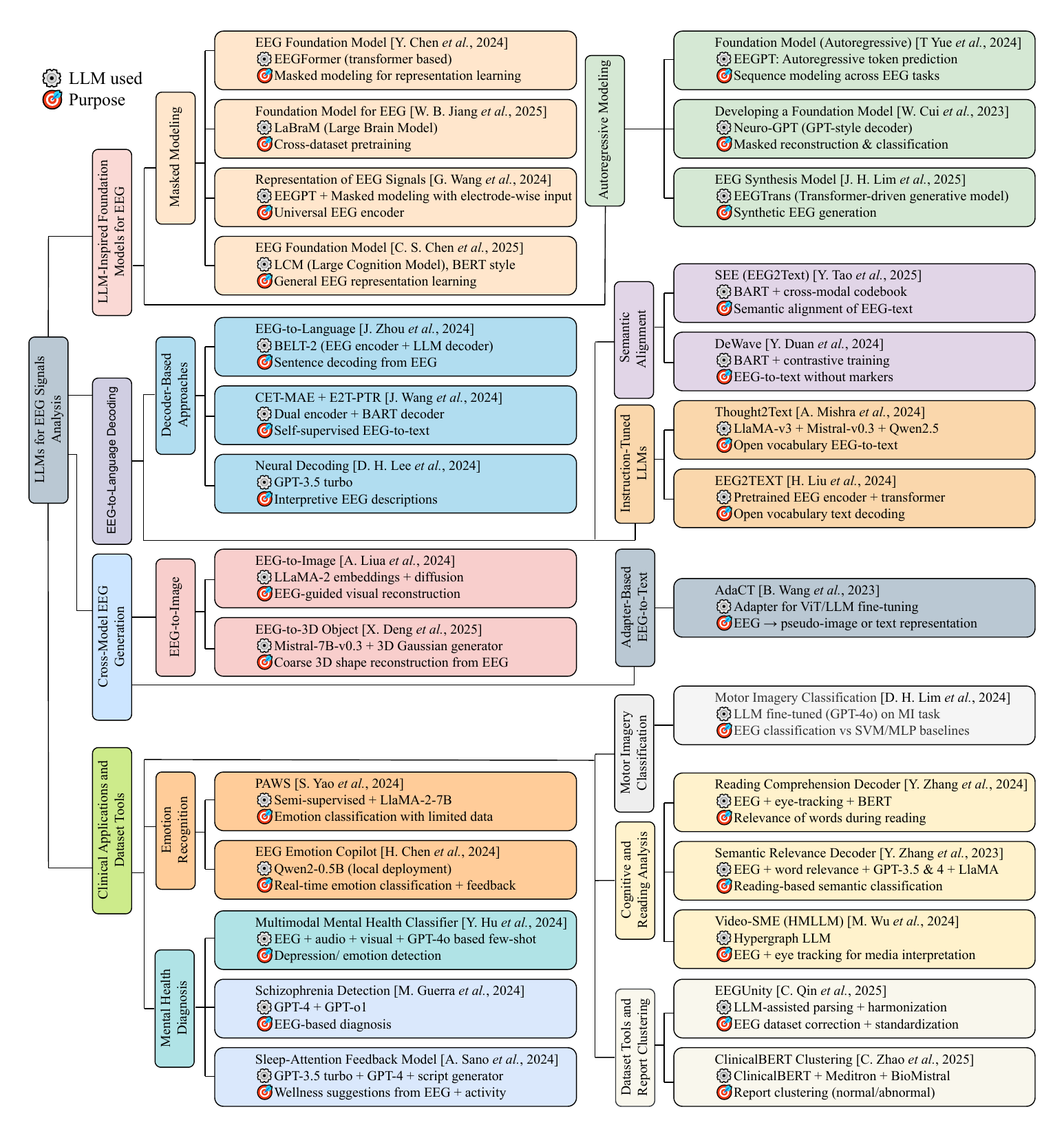}
        \caption{A hierarchical taxonomy of LLM-based EEG analysis, structured according to primary application domains. Each entry provides the model or paper name, approximate publication year, type of large language model, purpose, and task.}
    \label{hierarchy}
\end{figure*}

To illustrate the proposed taxonomy, this section reviews recent studies that integrate LLMs into EEG signal processing across various tasks, including emotion recognition, motor imagery, EEG-to-text translation, and clinical reporting. Despite differences in architecture and data requirements, these works demonstrate how language models enhance EEG data analysis and generation tasks. TABLE \ref{literature} provides a structured overview, summarizing each study by model or paper name, the integration of these models, their role (e.g., decoding, alignment, generation), and the broader task category. Figures \ref{llm_family} and \ref{hierarchy} further visualize the models and their applications across these tasks.

\begin{table*}[ht!]

\caption{This table provides a comprehensive overview of recent studies that integrate Large Language Models (LLMs) with Electroencephalography (EEG) signals across multiple domains. The table reports the model name, the use of LLMs and EEG data, the model's primary role, and the broader classification of each study.}
\centering
\resizebox{1\textwidth}{!}{%
\begin{tabular}{p{2.3cm}|p{0.8cm}|p{0.5cm}|p{6.5cm}|p{4cm}}
\hline
\textbf{Model Paper} & \textbf{LLMs Used} & \textbf{EEG Used} & \textbf{LLM Purpose} & \textbf{Category} \\
\hline
PAWS \cite{entry01} & Yes & Yes & Enhancing feature representation for semi-supervised emotion recognition for EEG data & Emotion Recognition with LLM \\
\hline
BELT-2 \cite{entry02} & Yes & Yes & Uses prefix tuning to align EEG embeddings with GPT for EEG-to-text generation & LLM-Based EEG-to-Language Decoding \\
\hline
GPT-4o \cite{entry03} & Yes & Yes & Fine-tunes LLM for motor imagery and compares it to traditional classifiers & Cognitive and Motor State Classification \\
\hline
BART+GPT-4 \cite{entry04} & Yes & Yes & Subject-dependent representation for open-vocabulary sentence generation and refinement & EEG-to-Language Decoding with Hybrid LLM Architecture \\
\hline
BART \cite{entry05} & Yes & Yes & Quantized variational encoder and contrastive training to align EEG tokens with BART & EEG-to-Text Generation via Discrete Contrastive Alignment \\
\hline
Qwen2-0.5B \cite{entry06} & Yes & Yes & 	Real-time emotion recognition and generates diagnostic feedback with prompt tuning & Emotion Recognition and Clinical Text Generation \\
\hline
BERT \cite{entry07} & Implicit & Yes & 	Contrastive learning of shared EEG-text embeddings for few-shot and zero-shot classification & Multimodal EEG Text Alignment with Contrastive Learning \\
\hline
EEG-GPT \cite{entry08} & Yes & Yes & Few-shot interpretable abnormality classification & Interpretable EEG Classification \\
\hline
ELM \cite{entry09} & Yes & Yes & EEG+text for pathology detection with multimodals & EEG for Clinical Pathology \\
\hline
EEGUnity \cite{entry10} & Yes & Yes & Uses an LLM Boost module for metadata harmonization, structure inference, and label standardization across EEG datasets	& EEG Dataset Management and Preprocessing with LLM Assistance\\
\hline
BART \cite{entry11} & Yes & Yes & Contrastive EEG-text learning + BART decoder & EEG-to-Language Decoding \\
\hline
GPT-3.5 \cite{entry12} & Yes & Yes & Interpretive neural decoding using GPT & Neural Decoding \\
\hline
GPT-3.5turbo \cite{entry13} & Yes & Yes & Wellness feedback, guided imagery via LLM & Behavioral Analysis \\
\hline
GPT-4o \cite{entry14} & Yes & Yes & Multimodal Depression/emotion prediction  & Mental Health Assessment \\
\hline
GPT-(4, o1)\cite{entry15} & Yes & Yes & Schizophrenia Detection through EEG Analysis  & Mental Health Diagnosis \\
\hline
LLaMA-2 \cite{entry16} & Yes & Yes & Visual decoding with semantic alignment & EEG-to-Image Visual Decoding \\
\hline
BERT \cite{entry17} & Yes & Yes & EEG+eye-tracking comprehension prediction & EEG Reading Analysis \\
\hline
HMLLM \cite{entry18} & Yes & Yes & Video content analysis with EEG + eye-tracking & Multimodal Cognitive EEG \\
\hline
GPT+LlaMA \cite{entry19} & Yes & Yes & Word-level semantic relevance prediction during reading using EEG and eye-tracking data & LLM-Guided EEG Semantic Decoding \\
\hline
LaBraM \cite{entry20} & LLM-style & Yes & Channel-wise patches, trains a vector-quantized transformer model for cross-dataset generalization & LLM-Inspired Foundation Models for EEG \\
\hline
LCM \cite{entry21} & LLM-style & Yes & Spectral-temporal attention mechanisms, generalization across multiple EEG datasets and tasks & LLM-Inspired Foundation Models for EEG \\
\hline
EEGPT \cite{entry22} & LLM-style & Yes & Masked spatio-temporal modeling with electrode-wise inputs, generalizable EEG representation & LLM-Inspired Foundation Models for EEG \\
\hline
AdaCT \cite{entry23} & Yes & Yes & EEG into pseudo-image or text form for fine-tuning pretrained vision and language transformers & Adapter-Based LLM Transfer Learning for EEG \\
\hline
ClinicalBERT \cite{entry24} & Yes & Text & EEG Report clustering with clinical LLM & EEG Report Analysis \\
\hline
Mistral-7B \cite{entry25} & Yes & Yes & Gaussian-based generative 3D object modeling & EEG-to-3D Object construction \\
\hline
Neuro-GPT \cite{entry26} & Yes & Yes & Masked segment reconstruction and generalization in low-data motor imagery tasks & Hybrid EEG Foundation Architectures \\
\hline
Transformer  \cite{entry27} & Yes & Yes & Open vocabulary EEG-to-text via pretraining & EEG-to-Text Generation \\
\hline
BART \cite{entry28} & Yes & Yes & Cross-modal codebook + semantic alignment  & EEG-to-Text \\
\hline
Multimodals \cite{entry29} & Yes & Yes & Instruction-tuned LLaMA, Mistral, and Qwen used & EEG-to-Text \\
\hline
EEGFormer  \cite{entry30} & LLM-style & Yes & Self-supervised transformer for universal EEG representations & LLM-Inspired Foundation Models \\
\hline
EEGTrans \cite{entry31} & LLM-style & Yes & Autoregressive EEG signal generator using transformer decoder & LLM-Inspired Generative Models for Synthetic EEG \\
\hline
EEGPT \cite{entry32} & LLM-style & Yes & Next-signal prediction with 1.1B model for multitask EEG & LLM-Style Foundation Models \\
\hline
\end{tabular}
\label{literature}
}
\end{table*}

\begin{table*}[ht]
\centering
\caption{EEG Datasets used in the LLM-Based Studies}
\resizebox{1\textwidth}{!}{%
\begin{tabular}{p{5cm}|p{4cm}|p{6cm}}
\hline
\textbf{Datasets} & \textbf{Papers} & \textbf{Purpose / Task} \\
\hline
SEED \cite{seed}, SEED-IV \cite{seed_iv} & \cite{entry01} (PAWS) & Emotion recognition  \\
\hline
ZuCo \cite{ZuCo} & \cite{entry02} (BELT-2), \cite{entry05} (DeWave), \cite{entry11} (CET-MAE), \cite{entry27} (EEG2TEXT), \cite{entry28} (SEE), \cite{entry29} (Thought2Text) & EEG-to-text translation, semantic relevance decoding  \\
\hline
ImageNet-EEG \cite{eeg_image01, eeg_image02} & \cite{entry16} (EEG-to-Image), \cite{entry25} (EEG-to-3D Generation) & Cross-modal visual and 3D generation \\
\hline
LMSU ScZ \cite{scz} & \cite{entry15} (LLM for Schizophrenia Detection) & Schizophrenia classification \\
\hline
Sleep-EDF \cite{sleep} & \cite{entry13} (Sleep-Attention Feedback Model), \cite{entry23} (AdaCT) & Sleep staging, EEG + activity data for wellness analysis \\
\hline
BCI competition III \cite{bci} & \cite{entry03} (Motor Imagery tasks) & Classification of non-invasive EEG signals during
Motor Imagery tasks using a Large Language Model \\
\hline
The Temple University Hospital EEG corpus (TUEG) \cite{temple} & \cite{entry07} (Learning EEG representations), \cite{entry08} (EEG-GPT: MODELS FOR EEG CLASSIFICATION), \cite{entry30} (EEGFormer: Foundation Model) & Learning EEG representations from natural language descriptions \\
\hline

TUEG \cite{temple}, TUAB \cite{TUAB}, NMT \cite{NMT}, TUSZ \cite{TUSZ} & \cite{entry09} (Learning EEG representations), \cite{entry08} (EEG-Language modeling) & EEG-language modeling for pathology detection \\
\hline

ZuCo 1.0 \cite{ZuCo01} & \cite{entry04} (EEG-to-Text), \cite{entry17, entry19} (Reading comprehension) & EET-to-text and reading analysis \\
\hline

SRI-ADV \cite{entry18} & \cite{entry18} (Hypergraph Multi-modal) & EEG and Eye-tracking Modalities to Evaluate Heterogeneous Responses for Video Understanding \\
\hline

TUAB, TUEV \cite{temple} & \cite{entry20} (Learning generic representation) & Large Brain Model for learning generic representation with tremendous EEG data in BCI \\
\hline

PhysioMI \cite{PhysioNet}, TSU \cite{tsu}, SEED \cite{seed}, BCIC-2A \cite{bci4}, BCIC-2B  \cite{bci2b} & \cite{entry21} (EEG Foundation Model) & Large Cognition Model: Towards Pretrained
Electroencephalography (EEG) Foundation Model \\
\hline

PhysioMI \cite{PhysioNet}, HGD \cite{hgd}, TSU \cite{tsu}, SEED \cite{seed}, M3CV \cite{m3cv}, BCIC-2A \cite{bci4}, BCIC-2B  \cite{bci2b}, Sleep-EDFx \cite{sleep}, KaggleERN \cite{kaggle}, PhysioP300 \cite{PhysioP300}, TUAB, TUEV \cite{temple}  & \cite{entry22} (Reliable Representation of EEG signals) & Pretrained Transformer for Universal and Reliable Representation of EEG Signals \\
\hline

BCI  Competition \cite{bci4}, HGD \cite{hgd}  & \cite{entry31} (Generative models for EEG synthesis) & EEGTRANS: Transformer-driven generative models for EEG synthesis \\
\hline

DEAP \cite{deap}, FACED \cite{faced}, SEED-IV \cite{seed_iv}, SEED-V \cite{seed_v}, MIBCI \cite{mibci}, EEGMat \cite{EEGMat}, STEW \cite{STEW}, HMC \cite{hmc}, IMG \cite{entry32}, SPE \cite{SPE}  & \cite{entry32} (Foundation model) & Unleashing the potential of EEG
generalist foundation model by autoregressive pre-training\\
\hline

\end{tabular}
\label{dataset_table}
}
\end{table*}

\subsection{LLM-Inspired Foundation Models for EEG}
Foundation models are trained on large-scale datasets using self-supervised learning objectives, such as masked modeling and autoregressive prediction. These models are designed to learn general-purpose, transferable representations that can support a variety of downstream tasks, including emotion recognition, abnormality detection, and cognitive state classification.

\subsubsection{Masked Pretraining (BERT-Style)}
Inspired by BERT's masked language modeling strategy, several EEG foundation models adopt similar self-supervised approaches by learning to reconstruct masked segments of input signals. EEGFormer applies masked token prediction over spatiotemporal patches using a transformer architecture \cite{entry30}, while LaBraM \cite{entry20} segments signals into channel-wise patches and employs a vector-quantized tokenizer to enable generalization across datasets. EEGPT \cite{entry22} introduces electrode-wise masked modeling, and LCM \cite{entry21} enhances this framework by incorporating both temporal and spectral attention during large-scale pretraining.

\subsubsection{Autoregressive Pretraining (GPT-style)}
Inspired by GPT's autoregressive modeling, which involves predicting the next token in a sequence, several models have adopted next-step prediction frameworks to capture temporal dependencies in brain signals. EEGPT includes an autoregressive variant that discretizes inputs and predicts future tokenized segments sequentially \cite{entry32}. Similarly, Neuro-GPT \cite{entry26} combines a pretrained EEG encoder with a GPT-style decoder for masked segment reconstruction, showing strong performance in low-data scenarios such as motor imagery classification. These models are particularly effective in tasks requiring temporal reasoning, including sequence prediction and classification under limited supervision.

Generating synthetic data is increasingly used to address challenges such as data scarcity, domain generalization, and class imbalance. While GAN-based methods have been popular \cite{gan01, gan02, gan03}, they often suffer from training instability and limited temporal precision. Recent transformer-based approaches, inspired by LLMs, provide a more stable alternative \cite{gan04, gan05}. EEGTrans \cite{entry31} applies a GPT-style transformer to generate discrete EEG sequences by first compressing raw signals with a quantized autoencoder, then learning temporal patterns autoregressively. This method produces high-fidelity synthetic EEG that preserves spectral and sequential characteristics, and has been shown to improve motor imagery classification performance while outperforming GAN-based baselines in signal quality and downstream accuracy.

\subsection{EEG-to-Language Decoding}
A particularly promising application of LLMs in EEG analysis is the decoding of brain activity into natural language. Recent models achieve this by coupling EEG encoders with pretrained language decoders such as BART and GPT, enabling both fixed and open-vocabulary text generation \cite{entry04}. Notable examples include BELT-2 \cite{entry02}, CET-MAE \cite{entry11}, and Thought2Text \cite{entry27}, which incorporate techniques like instruction tuning and semantic alignment to enhance the fluency and contextual relevance of the generated text. These models significantly advance EEG-to-text translation and offer strong potential for real-world brain-computer interface (BCI) systems that facilitate communication through neural signals.

\subsubsection{Decoder-Based Approaches}
Decoder-based EEG-to-text models typically pair a neural EEG encoder with a pretrained language decoder such as BART or GPT, allowing the model to generate natural language conditioned on brain activity. This modular design leverages the linguistic strength of LLMs while grounding them in neural representations. For instance, BELT-2 \cite{entry02} uses prefix tuning to connect an EEG encoder to a GPT-style decoder, aligning EEG embeddings with linguistic tokens. CET-MAE \cite{entry11} incorporates a dual-stream encoder and a BART decoder trained with multi-modal self-supervised objectives. Other approaches utilize general-purpose models like GPT-3.5 to generate interpretable descriptions of EEG activity \cite{entry12}.

\subsubsection{Instruction-Tuned LLMs}
Instruction-tuned large language models enable open-vocabulary decoding, allowing the generation of free-form text from neural activity instead of limiting output to predefined classes. Thought2Text \cite{entry29} fine-tunes instruction-optimized models such as LLaMA-V3, Mistral, and Qwen2.5 to produce descriptive text from EEG embeddings recorded during visual stimulus tasks. Its training pipeline involves three stages: EEG encoding, multimodal alignment using vision-language data, and EEG-to-text decoding. Similarly, EEG2TEXT \cite{entry27} introduces a brain-to-text framework that incorporates enhanced pretraining and a multi-view transformer to enable open-ended text generation.

\subsubsection{Semantic Alignment}
Recent models aim to improve EEG-to-text translation by aligning neural features with textual embeddings in a shared semantic space. DeWave \cite{entry05} employs a quantized variational encoder to generate discrete codes, which are aligned with language representations through contrastive training. SEE \cite{entry28} introduces a semantic matching module that maps signal features into a shared latent space with text using precomputed embeddings, while addressing false negatives during training. These approaches focus on reducing the modality gap between brain signals and language by learning semantically consistent representations, thereby enhancing the coherence of the generated text.

\subsection{Cross-Modal EEG Generation}
Recent work has explored translating neural signals into other modalities such as visual scenes and 3D objects. These approaches integrate brain representations into generative pipelines, such as diffusion models and Gaussian frameworks, guided by embeddings derived from large language models (LLMs) \cite{entry16, entry25}.

\subsubsection{EEG-to-Image Generation}
EEG-to-image generation utilizes LLaMA-2 as a semantic scaffold \cite{entry16}. In this framework, brain signals are first encoded into latent features aligned with image captions. LLaMA-2 then processes these to extract high-level semantic embeddings that guide the subsequent image synthesis. These embeddings condition a pretrained diffusion model that fuses them with signal-derived visual features to synthesize images capturing both the original stimulus's structural layout and semantic content. Similarly, EEG-to-3D object reconstruction has been achieved using language models as semantic intermediaries \cite{entry25}. In this method, a neural encoder extracts spatiotemporal features, which a fine-tuned LLM interprets to produce semantic embeddings. These guide a 3D Gaussian-based generator to reconstruct object geometry and layout. Although the generated models remain coarse, they reflect the conceptual form of the perceived or imagined object.

\subsubsection{Adapter-Based EEG-to-Text}
AdaCT \cite{entry23} introduces an adapter-based framework that enables flexible integration of EEG signals with both language and vision models. It features two plug-and-play modules: AdaCT-T, which converts short EEG segments into textual representations for fine-tuning language transformers; and AdaCT-I, which transforms longer sequences into 2D pseudo-images compatible with vision transformers. This dual-pathway design allows the system to adapt large language models and vision transformers (ViTs) for EEG decoding across modalities.

\subsection{Clinical Applications and Dataset Tools}
LLMs are increasingly being explored in clinical and affective computing contexts, enabling tasks such as emotion recognition, mental health monitoring, diagnostic report generation, and decision support. These models enhance the interpretability of brain signals and support more personalized, multimodal analysis of cognitive and emotional states. In parallel, they are also applied to EEG data management and augmentation, such as generating synthetic samples to improve model generalization, parsing and organizing large-scale datasets, and assisting with data cleaning and clustering. Semantic alignment techniques have further enabled the embedding of neural signals and clinical text into a shared space, supporting zero-shot and few-shot inference.

\subsubsection{Emotion Recognition}
Emotion recognition in low-resource settings remains a key challenge. To address this, \cite{entry01} proposes a semi-supervised framework that combines LLM-driven feature extraction with mixup-based data augmentation to boost performance when labeled data is scarce. Another system, EEG Emotion Copilot, employs a lightweight 0.5B-parameter LLM capable of real-time classification and personalized feedback generation \cite{entry06}. Designed for efficient on-device inference, it incorporates prompt learning, model pruning, and fine-tuning strategies. Beyond recognizing emotional states, the copilot also generates diagnostic summaries and treatment suggestions, supporting automated clinical documentation.

\subsubsection{Mental Health Diagnosis}
Mental health is a prime concern at present, and large language models are increasingly being explored for multimodal diagnostics. One line of work focuses on classifying depression and emotional states by integrating EEG signals with additional inputs such as facial expressions, audio, and behavioral data \cite{entry14}. These systems often employ zero-shot or few-shot prompting to generalize across mental health tasks without extensive retraining. In a related study, GPT-4 and GPT-01 were applied to classify schizophrenia from EEG recordings, producing interpretable outputs that align with established clinical patterns \cite{entry15}. Another model, the Sleep-Attention Feedback Model, fuses EEG and physical activity data to estimate attention states and sleep stages, using LLMs to generate personalized wellness guidance and adaptive imagery scripts \cite{entry13}. While performance may be constrained by limited labeled data, these efforts highlight the potential of LLMs in supporting both clinical and non-clinical applications for cognitive and emotional health.

\subsubsection{Motor Imagery Classification}
Motor imagery (MI) classification has emerged as a promising application of language models in EEG analysis. One study explores a fine-tuned LLM-based classifier that processes these signals recorded during MI tasks, demonstrating performance comparable to or exceeding traditional models such as Support Vector Machines, Random Forests, and Multi-Layer Perceptrons \cite{entry03}. In a related approach, Neuro-GPT \cite{entry26} introduces a hybrid framework that combines a pretrained EEG encoder with a GPT-style decoder for masked segment reconstruction. Evaluated under low-data conditions, the model shows strong potential for subject-generalization with minimal supervision, highlighting the advantages of transformer-based architectures in MI decoding.

\subsubsection{Cognitive and Reading Analysis}
Analyzing cognitive states during reading and video interpretation has become a key focus in LLM-guided EEG research. The Reading Comprehension Decoder \cite{entry17} combines EEG and eye-tracking data using an attention-based transformer encoder guided by word-level embeddings from a pretrained BERT model. It predicts the relevance of each word to a target inference keyword, achieving over 68\% accuracy across multiple subjects. Complementarily, the Semantic Relevance Decoder \cite{entry19} demonstrates that EEG signals alone can classify word-level understanding, highlighting the feasibility of decoding comprehension without additional modalities. Extending beyond reading, the Video-SME dataset \cite{entry18} investigates subjective video interpretation by capturing EEG and gaze data across diverse participants. A Hypergraph Multimodal LLM (HMLLM) is used to model semantic relationships among brain signals, gaze behavior, and video content, supporting deeper insights into cognitive and perceptual processing.

\subsubsection{Dataset Tools and Report Clustering}
Beyond decoding and classification, LLMs are increasingly used for managing, interpreting, and standardizing EEG datasets and clinical reports crucial tasks given the field’s challenges with data heterogeneity, inconsistent formatting, and weak alignment between signals and annotations. Language models pretrained on biomedical corpora offer strong capabilities for automating the parsing, structuring, and semantic organization of such data \cite{entry10, entry24}. A notable example is EEGUnity, an open-source framework that streamlines preprocessing across diverse sources \cite{entry10}. It features modules for raw file parsing, inconsistency correction, and batch processing, alongside an LLM Boost component that infers intelligent data structures, harmonizes metadata, and standardizes annotations. Evaluated on 25 public datasets, EEGUnity demonstrates robust performance in handling varied channel layouts, sampling rates, and labeling schemes, addressing fragmentation that often hinders large-scale analysis. Semantic clustering and joint learning approaches are reshaping how EEG reports and signals are interpreted using language models, like ClinicalBERT, Meditron-7B, and BioMistral have been used to group patients into diagnostic categories such as normal and abnormal \cite{entry24}, with fuzzy clustering methods (e.g., Fuzzy C-Means, Fuzzy J-Means) supporting soft label assignments to handle clinical ambiguity. For low-resource scenarios, EEG-GPT \cite{entry08} applies a few-shot transformer trained with just 2\% labeled data, achieving performance comparable to fully supervised models while offering interpretable, step-by-step reasoning. EEG-Language Model (ELM) \cite{entry09} further advances this space by jointly learning from EEG signals and their associated clinical reports using time-series cropping, text segmentation, and multiple instance learning. ELM supports both classification and retrieval, enabling bidirectional querying between neural data and medical documentation.

\section{Future Directions}\label{future}
While significant progress has been made in applying language models to EEG tasks, some challenges still need to be addressed to further enhance brain signal interpretation and enable real-world applications..

\begin{itemize}
\item \textbf{Real-Time and Low-Latency Inference:} Achieving real-time performance is essential for neurofeedback and assistive communication applications. This requires exploring model compression techniques- quantization, knowledge distillation, and dynamic attention to support efficient, low-latency streaming inference.

\item \textbf{Personalized and Adaptive Models:} Due to the high inter-subject variability in EEG signals, personalization is essential for achieving reliable performance. Techniques such as continual learning, meta-learning, and federated fine-tuning can enable models to adapt to individual users over time, thereby improving both accuracy and user trust.

\item \textbf{Multilingual and Multimodal Interfaces:} While EEG-based systems have traditionally focused on monolingual and unimodal processing, the emergence of multimodal large language models (LLMs) presents new opportunities to advance brain computer interaction. These include inner speech decoding from EEG, cross-lingual emotion recognition, and the integration of EEG with complementary biosignals such as eye tracking, facial expressions, and audio, enabling more robust and context-aware interpretation of brain activity.

\item \textbf{Interpretability and Clinical Validation:} Despite the strong performance of LLMs, they function as black boxes, limiting their adoption in clinical settings. To address this limitation, explainable AI techniques, including attention maps, saliency visualizations, and natural language rationales, should be emphasized to enhance transparency and support clinical decision-making.
\end{itemize}

\section{Conclusion}\label{conclusion}
This work presents a comprehensive survey of recent studies employing large language models (LLMs) in conjunction with EEG signal processing. The reviewed works are systematically categorized into four major domains: (1) LLM-inspired foundation models for EEG analysis, (2) EEG-to-language decoding, (3) cross-modal generation, including image and 3D object synthesis, and (4) clinical applications and dataset management tools, encompassing emotion recognition, mental health analysis, and clinical report interpretation. To unify these developments, a structured taxonomy is proposed, highlighting the application areas, utilized models, and the specific roles of these models. Furthermore, various adaptation strategies are examined, including fine-tuning, zero-shot learning, and few-shot learning. Finally, the survey discussed current limitations and outlined future directions.

\ifCLASSOPTIONcaptionsoff
  \newpage
\fi

\bibliographystyle{IEEEtran}
\bibliography{ref}

\begin{thebibliography}{100}
\providecommand{\url}[1]{#1}
\csname url@samestyle\endcsname
\providecommand{\newblock}{\relax}
\providecommand{\bibinfo}[2]{#2}
\providecommand{\BIBentrySTDinterwordspacing}{\spaceskip=0pt\relax}
\providecommand{\BIBentryALTinterwordstretchfactor}{4}
\providecommand{\BIBentryALTinterwordspacing}{\spaceskip=\fontdimen2\font plus
\BIBentryALTinterwordstretchfactor\fontdimen3\font minus \fontdimen4\font\relax}
\providecommand{\BIBforeignlanguage}[2]{{%
\expandafter\ifx\csname l@#1\endcsname\relax
\typeout{** WARNING: IEEEtran.bst: No hyphenation pattern has been}%
\typeout{** loaded for the language `#1'. Using the pattern for}%
\typeout{** the default language instead.}%
\else
\language=\csname l@#1\endcsname
\fi
#2}}
\providecommand{\BIBdecl}{\relax}
\BIBdecl

\bibitem{bert04}
\BIBentryALTinterwordspacing
B.~Ghojogh and A.~Ghodsi, ``{Attention Mechanism, Transformers, BERT, and GPT: Tutorial and Survey},'' Dec. 2020, working paper or preprint. [Online]. Available: \url{https://hal.science/hal-04637647}
\BIBentrySTDinterwordspacing

\bibitem{bert03}
M.~V. Koroteev, ``Bert: a review of applications in natural language processing and understanding,'' \emph{arXiv preprint arXiv:2103.11943}, 2021.

\bibitem{bert02}
\BIBentryALTinterwordspacing
G.~Jawahar, B.~Sagot, and D.~Seddah, ``What does {BERT} learn about the structure of language?'' in \emph{Proceedings of the 57th Annual Meeting of the Association for Computational Linguistics}, A.~Korhonen, D.~Traum, and L.~M{\`a}rquez, Eds.\hskip 1em plus 0.5em minus 0.4em\relax Florence, Italy: Association for Computational Linguistics, Jul. 2019, pp. 3651--3657. [Online]. Available: \url{https://aclanthology.org/P19-1356/}
\BIBentrySTDinterwordspacing

\bibitem{bci01}
L.~F. Nicolas-Alonso and J.~Gomez-Gil, ``Brain computer interfaces, a review,'' \emph{Sensors}, vol.~12, no.~2, pp. 1211--1279, Jan 2012.

\bibitem{bci02}
J.~Wolpaw, N.~Birbaumer, W.~Heetderks, D.~McFarland, P.~Peckham, G.~Schalk, E.~Donchin, L.~Quatrano, C.~Robinson, and T.~Vaughan, ``Brain-computer interface technology: a review of the first international meeting,'' \emph{IEEE Transactions on Rehabilitation Engineering}, vol.~8, no.~2, pp. 164--173, 2000.

\bibitem{eeg01}
D.~P. Subha, P.~K. Joseph, U.~R. Acharya, and C.~M. Lim, ``Eeg signal analysis: a survey,'' \emph{Journal of Medical Systems}, vol.~34, no.~2, pp. 195--212, Apr 2010.

\bibitem{eeg03}
\BIBentryALTinterwordspacing
J.~S. Kumar and P.~Bhuvaneswari, ``Analysis of electroencephalography (eeg) signals and its categorization–a study,'' \emph{Procedia Engineering}, vol.~38, pp. 2525--2536, 2012, iNTERNATIONAL CONFERENCE ON MODELLING OPTIMIZATION AND COMPUTING. [Online]. Available: \url{https://www.sciencedirect.com/science/article/pii/S1877705812022114}
\BIBentrySTDinterwordspacing

\bibitem{eeg04}
D.~P. Subha, P.~K. Joseph, R.~Acharya~U, and C.~M. Lim, ``Eeg signal analysis: a survey,'' \emph{Journal of medical systems}, vol.~34, pp. 195--212, 2010.

\bibitem{entry29}
A.~Mishra, S.~Shukla, J.~Torres, J.~Gwizdka, and S.~Roychowdhury, ``Thought2text: Text generation from eeg signal using large language models (llms),'' \emph{arXiv preprint arXiv:2410.07507}, 2024.

\bibitem{entry16}
A.~Liu, H.~Jing, Y.~Liu, Y.~Ma, and N.~Zheng, ``Hidden states in llms improve eeg representation learning and visual decoding,'' vol. 392, pp. 2130--2137, 2024.

\bibitem{entry25}
X.~Deng, S.~Chen, J.~Zhou, and L.~Li, ``Mind2matter: Creating 3d models from eeg signals,'' \emph{arXiv preprint arXiv:2504.11936}, 2025.

\bibitem{entry14}
Y.~Hu, S.~Zhang, T.~Dang, H.~Jia, F.~D. Salim, W.~Hu, and A.~J. Quigley, ``Exploring large-scale language models to evaluate eeg-based multimodal data for mental health,'' in \emph{Companion of the 2024 on ACM International Joint Conference on Pervasive and Ubiquitous Computing}, 2024, pp. 412--417.

\bibitem{artifact01}
\BIBentryALTinterwordspacing
J.~A. Urigüen and B.~Garcia-Zapirain, ``Eeg artifact removal—state-of-the-art and guidelines,'' \emph{Journal of Neural Engineering}, vol.~12, no.~3, p. 031001, apr 2015. [Online]. Available: \url{https://dx.doi.org/10.1088/1741-2560/12/3/031001}
\BIBentrySTDinterwordspacing

\bibitem{artifact02}
M.~M.~N. Mannan, M.~A. Kamran, and M.~Y. Jeong, ``Identification and removal of physiological artifacts from electroencephalogram signals: A review,'' \emph{IEEE Access}, vol.~6, pp. 30\,630--30\,652, 2018.

\bibitem{artifact03}
\BIBentryALTinterwordspacing
X.~Jiang, G.-B. Bian, and Z.~Tian, ``Removal of artifacts from eeg signals: A review,'' \emph{Sensors}, vol.~19, no.~5, 2019. [Online]. Available: \url{https://www.mdpi.com/1424-8220/19/5/987}
\BIBentrySTDinterwordspacing

\bibitem{subject_variability01}
\BIBentryALTinterwordspacing
G.~Huang, Z.~Zhao, S.~Zhang, Z.~Hu, J.~Fan, M.~Fu, J.~Chen, Y.~Xiao, J.~Wang, and G.~Dan, ``Discrepancy between inter- and intra-subject variability in eeg-based motor imagery brain-computer interface: Evidence from multiple perspectives,'' \emph{Frontiers in Neuroscience}, vol. Volume 17 - 2023, 2023. [Online]. Available: \url{https://www.frontiersin.org/journals/neuroscience/articles/10.3389/fnins.2023.1122661}
\BIBentrySTDinterwordspacing

\bibitem{subject_variability02}
\BIBentryALTinterwordspacing
S.~Saha and M.~Baumert, ``Intra- and inter-subject variability in eeg-based sensorimotor brain computer interface: A review,'' \emph{Frontiers in Computational Neuroscience}, vol. Volume 13 - 2019, 2020. [Online]. Available: \url{https://www.frontiersin.org/journals/computational-neuroscience/articles/10.3389/fncom.2019.00087}
\BIBentrySTDinterwordspacing

\bibitem{llm01}
\BIBentryALTinterwordspacing
Y.~Chang, X.~Wang, J.~Wang, Y.~Wu, L.~Yang, K.~Zhu, H.~Chen, X.~Yi, C.~Wang, Y.~Wang, W.~Ye, Y.~Zhang, Y.~Chang, P.~S. Yu, Q.~Yang, and X.~Xie, ``A survey on evaluation of large language models,'' \emph{ACM Trans. Intell. Syst. Technol.}, vol.~15, no.~3, Mar. 2024. [Online]. Available: \url{https://doi.org/10.1145/3641289}
\BIBentrySTDinterwordspacing

\bibitem{attention}
A.~Vaswani, N.~Shazeer, N.~Parmar, J.~Uszkoreit, L.~Jones, A.~N. Gomez, L.~Kaiser, and I.~Polosukhin, ``Attention is all you need,'' p. 6000–6010, 2017.

\bibitem{llm03}
T.~Shen, R.~Jin, Y.~Huang, C.~Liu, W.~Dong, Z.~Guo, X.~Wu, Y.~Liu, and D.~Xiong, ``Large language model alignment: A survey,'' \emph{arXiv preprint arXiv:2309.15025}, 2023.

\bibitem{llm05}
Y.~Wang, W.~Zhong, L.~Li, F.~Mi, X.~Zeng, W.~Huang, L.~Shang, X.~Jiang, and Q.~Liu, ``Aligning large language models with human: A survey,'' \emph{arXiv preprint arXiv:2307.12966}, 2023.

\bibitem{boka04}
\BIBentryALTinterwordspacing
Z.~Wan, M.~Li, S.~Liu, J.~Huang, H.~Tan, and W.~Duan, ``Eegformer: A transformer–based brain activity classification method using eeg signal,'' \emph{Frontiers in Neuroscience}, vol. Volume 17 - 2023, 2023. [Online]. Available: \url{https://www.frontiersin.org/journals/neuroscience/articles/10.3389/fnins.2023.1148855}
\BIBentrySTDinterwordspacing

\bibitem{boka05}
Z.~Wang, Y.~Wang, C.~Hu, Z.~Yin, and Y.~Song, ``Transformers for eeg-based emotion recognition: A hierarchical spatial information learning model,'' \emph{IEEE Sensors Journal}, vol.~22, no.~5, pp. 4359--4368, 2022.

\bibitem{entry27}
H.~Liu, D.~Hajialigol, B.~Antony, A.~Han, and X.~Wang, ``Eeg2text: Open vocabulary eeg-to-text decoding with eeg pre-training and multi-view transformer,'' \emph{arXiv preprint arXiv:2405.02165}, 2024.

\bibitem{bci04}
U.~Chaudhary, N.~Birbaumer, and A.~Ramos-Murguialday, ``Brain–computer interfaces for communication and rehabilitation,'' \emph{Nature Reviews Neurology}, vol.~12, no.~9, pp. 513--525, 2016.

\bibitem{motor_imagery01}
\BIBentryALTinterwordspacing
M.~Hamedi, S.-H. Salleh, and A.~M. Noor, ``Electroencephalographic motor imagery brain connectivity analysis for bci: A review,'' \emph{Neural Computation}, vol.~28, no.~6, pp. 999--1041, 06 2016. [Online]. Available: \url{https://doi.org/10.1162/NECO\_a\_00838}
\BIBentrySTDinterwordspacing

\bibitem{motor_imagery02}
\BIBentryALTinterwordspacing
M.~Ahn and S.~C. Jun, ``Performance variation in motor imagery brain–computer interface: A brief review,'' \emph{Journal of Neuroscience Methods}, vol. 243, pp. 103--110, 2015. [Online]. Available: \url{https://www.sciencedirect.com/science/article/pii/S0165027015000400}
\BIBentrySTDinterwordspacing

\bibitem{SSVEP01}
\BIBentryALTinterwordspacing
F.-B. Vialatte, M.~Maurice, J.~Dauwels, and A.~Cichocki, ``Steady-state visually evoked potentials: Focus on essential paradigms and future perspectives,'' \emph{Progress in Neurobiology}, vol.~90, no.~4, pp. 418--438, 2010. [Online]. Available: \url{https://www.sciencedirect.com/science/article/pii/S0301008209001853}
\BIBentrySTDinterwordspacing

\bibitem{SSVEP02}
\BIBentryALTinterwordspacing
G.~R. Müller-Putz, R.~Scherer, C.~Brauneis, and G.~Pfurtscheller, ``Steady-state visual evoked potential (ssvep)-based communication: impact of harmonic frequency components,'' \emph{Journal of Neural Engineering}, vol.~2, no.~4, p. 123, oct 2005. [Online]. Available: \url{https://dx.doi.org/10.1088/1741-2560/2/4/008}
\BIBentrySTDinterwordspacing

\bibitem{llm06}
\BIBentryALTinterwordspacing
Z.~Liang, Y.~Xu, Y.~Hong, P.~Shang, Q.~Wang, Q.~Fu, and K.~Liu, ``A survey of multimodel large language models,'' in \emph{Proceedings of the 3rd International Conference on Computer, Artificial Intelligence and Control Engineering}, ser. CAICE '24.\hskip 1em plus 0.5em minus 0.4em\relax New York, NY, USA: Association for Computing Machinery, 2024, p. 405–409. [Online]. Available: \url{https://doi.org/10.1145/3672758.3672824}
\BIBentrySTDinterwordspacing

\bibitem{llm07}
\BIBentryALTinterwordspacing
N.~Parmar, A.~Vaswani, J.~Uszkoreit, L.~Kaiser, N.~Shazeer, and A.~Ku, ``Image transformer,'' \emph{CoRR}, vol. abs/1802.05751, 2018. [Online]. Available: \url{http://arxiv.org/abs/1802.05751}
\BIBentrySTDinterwordspacing

\bibitem{han2021pre}
\BIBentryALTinterwordspacing
X.~Han, Z.~Zhang, N.~Ding, Y.~Gu, X.~Liu, Y.~Huo, J.~Qiu, Y.~Yao, A.~Zhang, L.~Zhang, W.~Han, M.~Huang, Q.~Jin, Y.~Lan, Y.~Liu, Z.~Liu, Z.~Lu, X.~Qiu, R.~Song, J.~Tang, J.-R. Wen, J.~Yuan, W.~X. Zhao, and J.~Zhu, ``Pre-trained models: Past, present and future,'' \emph{AI Open}, vol.~2, pp. 225--250, 2021. [Online]. Available: \url{https://www.sciencedirect.com/science/article/pii/S2666651021000231}
\BIBentrySTDinterwordspacing

\bibitem{bert01}
\BIBentryALTinterwordspacing
J.~Devlin, M.~Chang, K.~Lee, and K.~Toutanova, ``{BERT:} pre-training of deep bidirectional transformers for language understanding,'' \emph{CoRR}, vol. abs/1810.04805, 2018. [Online]. Available: \url{http://arxiv.org/abs/1810.04805}
\BIBentrySTDinterwordspacing

\bibitem{alayrac2022flamingo}
J.-B. Alayrac \emph{et~al.}, ``Flamingo: a visual language model for few-shot learning,'' \emph{arXiv preprint arXiv:2204.14198}, 2022.

\bibitem{liu2023llava}
H.~Liu, C.~Zhang, Z.~Hu, Y.~Wang, Z.~Yang, J.~Wang, W.~Chen \emph{et~al.}, ``Visual instruction tuning,'' \emph{arXiv preprint arXiv:2304.08485}, 2023.

\bibitem{wei2021finetuned}
J.~Wei, M.~Bosma, V.~Y. Zhao, K.~Guu, A.~W. Yu, B.~Lester, N.~Du, A.~M. Dai, and Q.~V. Le, ``Finetuned language models are zero-shot learners,'' \emph{arXiv preprint arXiv:2109.01652}, 2021.

\bibitem{fine_tuning}
S.~Zhang, L.~Dong, X.~Li, S.~Zhang, X.~Sun, S.~Wang, J.~Li, R.~Hu, T.~Zhang, F.~Wu \emph{et~al.}, ``Instruction tuning for large language models: A survey,'' \emph{arXiv preprint arXiv:2308.10792}, 2023.

\bibitem{entry23}
B.~Wang, X.~Fu, Y.~Lan, L.~Zhang, W.~Zheng, and Y.~Xiang, ``Large transformers are better eeg learners,'' \emph{arXiv preprint arXiv:2308.11654}, 2023.

\bibitem{entry02}
J.~Zhou, Y.~Duan, F.~Chang, T.~Do, Y.-K. Wang, and C.-T. Lin, ``Belt-2: Bootstrapping eeg-to-language representation alignment for multi-task brain decoding,'' \emph{arXiv preprint arXiv:2409.00121}, 2024.

\bibitem{zeroshot01}
\BIBentryALTinterwordspacing
W.~Wang, V.~W. Zheng, H.~Yu, and C.~Miao, ``A survey of zero-shot learning: Settings, methods, and applications,'' \emph{ACM Trans. Intell. Syst. Technol.}, vol.~10, no.~2, Jan. 2019. [Online]. Available: \url{https://doi.org/10.1145/3293318}
\BIBentrySTDinterwordspacing

\bibitem{zeroshot02}
F.~Pourpanah, M.~Abdar, Y.~Luo, X.~Zhou, R.~Wang, C.~P. Lim, X.-Z. Wang, and Q.~M.~J. Wu, ``A review of generalized zero-shot learning methods,'' \emph{IEEE Transactions on Pattern Analysis and Machine Intelligence}, vol.~45, no.~4, pp. 4051--4070, 2023.

\bibitem{entry07}
\BIBentryALTinterwordspacing
T.~C. N'dir and R.~T. Schirrmeister, ``Eeg-clip : Learning eeg representations from natural language descriptions,'' 2025. [Online]. Available: \url{https://arxiv.org/abs/2503.16531}
\BIBentrySTDinterwordspacing

\bibitem{entry18}
\BIBentryALTinterwordspacing
M.~Wu, C.~Zhao, A.~Su, D.~Di, T.~Fu, D.~An, M.~He, Y.~Gao, M.~Ma, K.~Yan, and P.~Wang, ``Hypergraph multi-modal large language model: Exploiting eeg and eye-tracking modalities to evaluate heterogeneous responses for video understanding,'' 2024. [Online]. Available: \url{https://arxiv.org/abs/2407.08150}
\BIBentrySTDinterwordspacing

\bibitem{fewshot01}
\BIBentryALTinterwordspacing
Y.~Wang, Q.~Yao, J.~Kwok, and L.~M. Ni, ``Generalizing from a few examples: A survey on few-shot learning,'' 2020. [Online]. Available: \url{https://arxiv.org/abs/1904.05046}
\BIBentrySTDinterwordspacing

\bibitem{fewshot02}
\BIBentryALTinterwordspacing
Y.~Song, T.~Wang, S.~K. Mondal, and J.~P. Sahoo, ``A comprehensive survey of few-shot learning: Evolution, applications, challenges, and opportunities,'' 2022. [Online]. Available: \url{https://arxiv.org/abs/2205.06743}
\BIBentrySTDinterwordspacing

\bibitem{entry06}
\BIBentryALTinterwordspacing
H.~Chen, W.~Zeng, C.~Chen, L.~Cai, F.~Wang, Y.~Shi, L.~Wang, W.~Zhang, Y.~Li, H.~Yan, W.~T. Siok, and N.~Wang, ``Eeg emotion copilot: Optimizing lightweight llms for emotional eeg interpretation with assisted medical record generation,'' 2025. [Online]. Available: \url{https://arxiv.org/abs/2410.00166}
\BIBentrySTDinterwordspacing

\bibitem{entry01}
S.~Yao, L.~Liu, J.~Lu, D.~Wu, and Y.~Li, ``Advancing semi-supervised eeg emotion recognition through feature extraction with mixup and large language models,'' in \emph{2024 IEEE International Conference on Bioinformatics and Biomedicine (BIBM)}, 2024, pp. 2772--2779.

\bibitem{entry03}
D.-H. Lim, M.-J. Cho, and H.-H. Kim, ``Classification of non-invasive eeg signals during motor imagery tasks using a large language model,'' in \emph{2024 International Conference on Cyberworlds (CW)}, 2024, pp. 378--379.

\bibitem{entry04}
\BIBentryALTinterwordspacing
H.~Amrani, D.~Micucci, and P.~Napoletano, ``Deep representation learning for open vocabulary electroencephalography-to-text decoding,'' \emph{IEEE Journal of Biomedical and Health Informatics}, p. 1–12, 2024. [Online]. Available: \url{http://dx.doi.org/10.1109/JBHI.2024.3416066}
\BIBentrySTDinterwordspacing

\bibitem{entry05}
Y.~Duan, J.~Zhou, Z.~Wang, Y.-K. Wang, and C.-T. Lin, ``Dewave: Discrete eeg waves encoding for brain dynamics to text translation,'' \emph{arXiv preprint arXiv:2309.14030}, 2023.

\bibitem{entry08}
J.~W. Kim, A.~Alaa, and D.~Bernardo, ``Eeg-gpt: exploring capabilities of large language models for eeg classification and interpretation,'' \emph{arXiv preprint arXiv:2401.18006}, 2024.

\bibitem{entry09}
S.~Gijsen and K.~Ritter, ``Eeg-language modeling for pathology detection,'' \emph{arXiv preprint arXiv:2409.07480}, 2024.

\bibitem{entry10}
C.~Qin, R.~Yang, W.~You, Z.~Chen, L.~Zhu, M.~Huang, and Z.~Wang, ``Eegunity: Open-source tool in facilitating unified eeg datasets toward large-scale eeg model,'' \emph{IEEE Transactions on Neural Systems and Rehabilitation Engineering}, vol.~33, pp. 1653--1663, 2025.

\bibitem{entry11}
J.~Wang, Z.~Song, Z.~Ma, X.~Qiu, M.~Zhang, and Z.~Zhang, ``Enhancing eeg-to-text decoding through transferable representations from pre-trained contrastive eeg-text masked autoencoder,'' \emph{arXiv preprint arXiv:2402.17433}, 2024.

\bibitem{entry12}
D.~H. Lee and C.~K. Chung, ``Enhancing neural decoding with large language models: A gpt-based approach,'' in \emph{2024 12th International Winter Conference on Brain-Computer Interface (BCI)}, 2024, pp. 1--4.

\bibitem{entry13}
A.~Sano, J.~Amores, and M.~Czerwinski, ``Exploration of llms, eeg, and behavioral data to measure and support attention and sleep,'' \emph{arXiv preprint arXiv:2408.07822}, 2024.

\bibitem{entry15}
M.~Guerra, R.~Milanese, M.~Deodato, M.~G. Ciobanu, and F.~Fasano, ``Exploring the diagnostic potential of llms in schizophrenia detection through eeg analysis,'' in \emph{2024 IEEE International Conference on Bioinformatics and Biomedicine (BIBM)}, 2024, pp. 6812--6819.

\bibitem{entry17}
Y.~Zhang, S.~Yang, G.~Cauwenberghs, and T.-P. Jung, ``From word embedding to reading embedding using large language model, eeg and eye-tracking,'' in \emph{2024 46th Annual International Conference of the IEEE Engineering in Medicine and Biology Society (EMBC)}, 2024, pp. 1--4.

\bibitem{entry19}
Y.~Zhang, Q.~Li, S.~Nahata, T.~Jamal, S.-k. Cheng, G.~Cauwenberghs, and T.-P. Jung, ``Integrating llm, eeg, and eye-tracking biomarker analysis for word-level neural state classification in semantic inference reading comprehension,'' \emph{arXiv preprint arXiv:2309.15714}, 2023.

\bibitem{entry20}
W.-B. Jiang, L.-M. Zhao, and B.-L. Lu, ``Large brain model for learning generic representations with tremendous eeg data in bci,'' \emph{arXiv preprint arXiv:2405.18765}, 2024.

\bibitem{entry21}
C.-S. Chen, Y.-J. Chen, and A.~H.-W. Tsai, ``Large cognition model: Towards pretrained eeg foundation model,'' \emph{arXiv preprint arXiv:2502.17464}, 2025.

\bibitem{entry22}
\BIBentryALTinterwordspacing
G.~Wang, W.~Liu, Y.~He, C.~Xu, L.~Ma, and H.~Li, ``Eegpt: Pretrained transformer for universal and reliable representation of eeg signals,'' in \emph{Advances in Neural Information Processing Systems}, A.~Globerson, L.~Mackey, D.~Belgrave, A.~Fan, U.~Paquet, J.~Tomczak, and C.~Zhang, Eds., vol.~37.\hskip 1em plus 0.5em minus 0.4em\relax Curran Associates, Inc., 2024, pp. 39\,249--39\,280. [Online]. Available: \url{https://proceedings.neurips.cc/paper_files/paper/2024/file/4540d267eeec4e5dbd9dae9448f0b739-Paper-Conference.pdf}
\BIBentrySTDinterwordspacing

\bibitem{entry24}
C.~Zhao, Z.~Cook, L.~Murray, J.~Kesan, N.~Belacel, S.~Doesburg, G.~Medvedev, V.~Vakorin, and P.~Xi, ``Leveraging large language models and fuzzy clustering for eeg report analysis,'' in \emph{2024 IEEE SENSORS}, 2024, pp. 1--4.

\bibitem{entry26}
W.~Cui, W.~Jeong, P.~Th{\"o}lke, T.~Medani, K.~Jerbi, A.~A. Joshi, and R.~M. Leahy, ``Neuro-gpt: Developing a foundation model for eeg,'' \emph{arXiv preprint arXiv:2311.03764}, vol. 107, 2023.

\bibitem{entry28}
Y.~Tao, Y.~Liang, L.~Wang, Y.~Li, Q.~Yang, and H.~Zhang, ``See: Semantically aligned eeg-to-text translation,'' in \emph{ICASSP 2025 - 2025 IEEE International Conference on Acoustics, Speech and Signal Processing (ICASSP)}, 2025, pp. 1--5.

\bibitem{entry30}
Y.~Chen, K.~Ren, K.~Song, Y.~Wang, Y.~Wang, D.~Li, and L.~Qiu, ``Eegformer: Towards transferable and interpretable large-scale eeg foundation model,'' \emph{arXiv preprint arXiv:2401.10278}, 2024.

\bibitem{entry31}
\BIBentryALTinterwordspacing
J.-H. Lim and P.-C. Kuo, ``{EEGT}rans: Transformer-driven generative models for {EEG} synthesis,'' 2025. [Online]. Available: \url{https://openreview.net/forum?id=ydw2l8zgUB}
\BIBentrySTDinterwordspacing

\bibitem{entry32}
T.~Yue, S.~Xue, X.~Gao, Y.~Tang, L.~Guo, J.~Jiang, and J.~Liu, ``Eegpt: Unleashing the potential of eeg generalist foundation model by autoregressive pre-training,'' \emph{arXiv preprint arXiv:2410.19779}, 2024.

\bibitem{seed}
W.-L. Zheng and B.-L. Lu, ``Investigating critical frequency bands and channels for eeg-based emotion recognition with deep neural networks,'' \emph{IEEE Transactions on Autonomous Mental Development}, vol.~7, no.~3, pp. 162--175, 2015.

\bibitem{seed_iv}
W.-L. Zheng, W.~Liu, Y.~Lu, B.-L. Lu, and A.~Cichocki, ``Emotionmeter: A multimodal framework for recognizing human emotions,'' \emph{IEEE Transactions on Cybernetics}, vol.~49, no.~3, pp. 1110--1122, 2019.

\bibitem{ZuCo}
\BIBentryALTinterwordspacing
N.~Hollenstein, M.~Troendle, C.~Zhang, and N.~Langer, ``\BIBforeignlanguage{eng}{{Z}u{C}o 2.0: A dataset of physiological recordings during natural reading and annotation},'' in \emph{\BIBforeignlanguage{eng}{Proceedings of the Twelfth Language Resources and Evaluation Conference}}, N.~Calzolari, F.~B{\'e}chet, P.~Blache, K.~Choukri, C.~Cieri, T.~Declerck, S.~Goggi, H.~Isahara, B.~Maegaard, J.~Mariani, H.~Mazo, A.~Moreno, J.~Odijk, and S.~Piperidis, Eds.\hskip 1em plus 0.5em minus 0.4em\relax Marseille, France: European Language Resources Association, May 2020, pp. 138--146. [Online]. Available: \url{https://aclanthology.org/2020.lrec-1.18/}
\BIBentrySTDinterwordspacing

\bibitem{eeg_image01}
C.~Spampinato, S.~Palazzo, I.~Kavasidis, D.~Giordano, N.~Souly, and M.~Shah, ``Deep learning human mind for automated visual classification,'' in \emph{2017 IEEE Conference on Computer Vision and Pattern Recognition (CVPR)}, 2017, pp. 4503--4511.

\bibitem{eeg_image02}
S.~Palazzo, C.~Spampinato, I.~Kavasidis, D.~Giordano, J.~Schmidt, and M.~Shah, ``Decoding brain representations by multimodal learning of neural activity and visual features,'' \emph{IEEE Transactions on Pattern Analysis and Machine Intelligence}, vol.~43, no.~11, pp. 3833--3849, 2021.

\bibitem{scz}
\BIBentryALTinterwordspacing
S.~V. Borisov, A.~Y. Kaplan, N.~L. Gorbachevskaya, and I.~A. Kozlova, ``Analysis of eeg structural synchrony in adolescents with schizophrenic disorders,'' \emph{Human Physiology}, vol.~31, no.~3, pp. 255--261, 2005. [Online]. Available: \url{https://doi.org/10.1007/s10747-005-0042-z}
\BIBentrySTDinterwordspacing

\bibitem{sleep}
B.~Kemp, A.~Zwinderman, B.~Tuk, H.~Kamphuisen, and J.~Oberye, ``Analysis of a sleep-dependent neuronal feedback loop: the slow-wave microcontinuity of the eeg,'' \emph{IEEE Transactions on Biomedical Engineering}, vol.~47, no.~9, pp. 1185--1194, 2000.

\bibitem{bci}
B.~Blankertz, K.-R. Muller, D.~Krusienski, G.~Schalk, J.~Wolpaw, A.~Schlogl, G.~Pfurtscheller, J.~Millan, M.~Schroder, and N.~Birbaumer, ``The bci competition iii: validating alternative approaches to actual bci problems,'' \emph{IEEE Transactions on Neural Systems and Rehabilitation Engineering}, vol.~14, no.~2, pp. 153--159, 2006.

\bibitem{temple}
\BIBentryALTinterwordspacing
I.~Obeid and J.~Picone, ``The temple university hospital eeg data corpus,'' \emph{Frontiers in Neuroscience}, vol. Volume 10 - 2016, 2016. [Online]. Available: \url{https://www.frontiersin.org/journals/neuroscience/articles/10.3389/fnins.2016.00196}
\BIBentrySTDinterwordspacing

\bibitem{TUAB}
S.~López, G.~Suarez, D.~Jungreis, I.~Obeid, and J.~Picone, ``Automated identification of abnormal adult eegs,'' in \emph{2015 IEEE Signal Processing in Medicine and Biology Symposium (SPMB)}, 2015, p. 10.1109/SPMB.2015.7405423.

\bibitem{NMT}
\BIBentryALTinterwordspacing
H.~A. Khan, R.~Ul~Ain, A.~M. Kamboh, H.~T. Butt, S.~Shafait, W.~Alamgir, D.~Stricker, and F.~Shafait, ``The nmt scalp eeg dataset: An open-source annotated dataset of healthy and pathological eeg recordings for predictive modeling,'' \emph{Frontiers in Neuroscience}, vol. Volume 15 - 2021, 2022. [Online]. Available: \url{https://www.frontiersin.org/journals/neuroscience/articles/10.3389/fnins.2021.755817}
\BIBentrySTDinterwordspacing

\bibitem{TUSZ}
V.~Shah, E.~von Weltin, S.~Lopez, J.~R. McHugh, L.~Veloso, M.~Golmohammadi, I.~Obeid, and J.~Picone, ``The temple university hospital seizure detection corpus,'' \emph{Frontiers in Neuroinformatics}, vol.~12, p.~83, Nov 2018.

\bibitem{ZuCo01}
N.~Hollenstein, J.~Rotsztejn, M.~Troendle, A.~Pedroni, C.~Zhang, and N.~Langer, ``Zuco, a simultaneous eeg and eye-tracking resource for natural sentence reading,'' \emph{Scientific Data}, vol.~5, p. 180291, Dec 2018.

\bibitem{PhysioNet}
A.~L. Goldberger, L.~A.~N. Amaral, L.~Glass, J.~M. Hausdorff, P.~C. Ivanov, R.~G. Mark, J.~E. Mietus, G.~B. Moody, C.-K. Peng, and H.~E. Stanley, ``Physiobank, physiotoolkit, and physionet: components of a new research resource for complex physiologic signals,'' \emph{Circulation}, vol. 101, no.~23, pp. e215--e220, Jun 2000.

\bibitem{tsu}
Y.~Wang, X.~Chen, X.~Gao, and S.~Gao, ``A benchmark dataset for ssvep-based brain-computer interfaces,'' \emph{IEEE Transactions on Neural Systems and Rehabilitation Engineering}, vol.~25, no.~10, pp. 1746--1752, Oct 2017, epub 2016 Nov 10.

\bibitem{bci4}
\BIBentryALTinterwordspacing
M.~Tangermann, K.-R. Müller, A.~Aertsen, N.~Birbaumer, C.~Braun, C.~Brunner, R.~Leeb, C.~Mehring, K.~J. Miller, G.~Mueller-Putz, G.~Nolte, G.~Pfurtscheller, H.~Preissl, G.~Schalk, A.~Schlögl, C.~Vidaurre, S.~Waldert, and B.~Blankertz, ``Review of the bci competition iv,'' \emph{Frontiers in Neuroscience}, vol. Volume 6 - 2012, 2012. [Online]. Available: \url{https://www.frontiersin.org/journals/neuroscience/articles/10.3389/fnins.2012.00055}
\BIBentrySTDinterwordspacing

\bibitem{bci2b}
\BIBentryALTinterwordspacing
D.~Steyrl, R.~Scherer, J.~Faller, and G.~R. Müller-Putz, ``Random forests in non-invasive sensorimotor rhythm brain-computer interfaces: a practical and convenient non-linear classifier,'' \emph{Biomedical Engineering / Biomedizinische Technik}, vol.~61, no.~1, pp. 77--86, 2016. [Online]. Available: \url{https://doi.org/10.1515/bmt-2014-0117}
\BIBentrySTDinterwordspacing

\bibitem{hgd}
\BIBentryALTinterwordspacing
R.~T. Schirrmeister, J.~T. Springenberg, L.~D.~J. Fiederer, M.~Glasstetter, K.~Eggensperger, M.~Tangermann, F.~Hutter, W.~Burgard, and T.~Ball, ``Deep learning with convolutional neural networks for eeg decoding and visualization,'' \emph{Human Brain Mapping}, vol.~38, no.~11, pp. 5391--5420, 2017. [Online]. Available: \url{https://onlinelibrary.wiley.com/doi/abs/10.1002/hbm.23730}
\BIBentrySTDinterwordspacing

\bibitem{m3cv}
\BIBentryALTinterwordspacing
G.~Huang, Z.~Hu, W.~Chen, Z.~Liang, L.~Li, L.~Zhang, and Z.~Zhang, ``M3cv:a multi-subject, multi-session, and multi-task database for eeg-based biometrics challenge,'' \emph{bioRxiv}, 2022. [Online]. Available: \url{https://www.biorxiv.org/content/early/2022/07/03/2022.06.28.497624}
\BIBentrySTDinterwordspacing

\bibitem{kaggle}
\BIBentryALTinterwordspacing
P.~Margaux, M.~Emmanuel, D.~Sébastien, B.~Olivier, and M.~Jérémie, ``Objective and subjective evaluation of online error correction during p300-based spelling,'' \emph{Advances in Human-Computer Interaction}, vol. 2012, no.~1, p. 578295, 2012. [Online]. Available: \url{https://onlinelibrary.wiley.com/doi/abs/10.1155/2012/578295}
\BIBentrySTDinterwordspacing

\bibitem{PhysioP300}
\BIBentryALTinterwordspacing
A.~L. Goldberger, L.~A.~N. Amaral, L.~Glass, J.~M. Hausdorff, P.~C. Ivanov, R.~G. Mark, J.~E. Mietus, G.~B. Moody, C.-K. Peng, and H.~E. Stanley, ``Physiobank, physiotoolkit, and physionet,'' \emph{Circulation}, vol. 101, no.~23, pp. e215--e220, 2000. [Online]. Available: \url{https://www.ahajournals.org/doi/abs/10.1161/01.CIR.101.23.e215}
\BIBentrySTDinterwordspacing

\bibitem{deap}
S.~Koelstra, C.~Muhl, M.~Soleymani, J.-S. Lee, A.~Yazdani, T.~Ebrahimi, T.~Pun, A.~Nijholt, and I.~Patras, ``Deap: A database for emotion analysis ;using physiological signals,'' \emph{IEEE Transactions on Affective Computing}, vol.~3, no.~1, pp. 18--31, 2012.

\bibitem{faced}
\BIBentryALTinterwordspacing
J.~Chen, X.~Wang, C.~Huang, X.~Hu, X.~Shen, and D.~Zhang, ``A large finer-grained affective computing eeg dataset,'' \emph{Scientific Data}, vol.~10, no.~1, p. 740, 2023. [Online]. Available: \url{https://doi.org/10.1038/s41597-023-02650-w}
\BIBentrySTDinterwordspacing

\bibitem{seed_v}
W.~Liu, J.-L. Qiu, W.-L. Zheng, and B.-L. Lu, ``Comparing recognition performance and robustness of multimodal deep learning models for multimodal emotion recognition,'' \emph{IEEE Transactions on Cognitive and Developmental Systems}, vol.~14, no.~2, pp. 715--729, 2022.

\bibitem{mibci}
\BIBentryALTinterwordspacing
H.~Cho, M.~Ahn, S.~Ahn, M.~Kwon, and S.~C. Jun, ``Eeg datasets for motor imagery brain–computer interface,'' \emph{GigaScience}, vol.~6, no.~7, p. gix034, 05 2017. [Online]. Available: \url{https://doi.org/10.1093/gigascience/gix034}
\BIBentrySTDinterwordspacing

\bibitem{EEGMat}
\BIBentryALTinterwordspacing
I.~Zyma, S.~Tukaev, I.~Seleznov, K.~Kiyono, A.~Popov, M.~Chernykh, and O.~Shpenkov, ``Electroencephalograms during mental arithmetic task performance,'' \emph{Data}, vol.~4, no.~1, 2019. [Online]. Available: \url{https://www.mdpi.com/2306-5729/4/1/14}
\BIBentrySTDinterwordspacing

\bibitem{STEW}
W.~L. Lim, O.~Sourina, and L.~P. Wang, ``Stew: Simultaneous task eeg workload data set,'' \emph{IEEE Transactions on Neural Systems and Rehabilitation Engineering}, vol.~26, no.~11, pp. 2106--2114, 2018.

\bibitem{hmc}
D.~Alvarez-Estevez and R.~Rijsman, ``Haaglanden medisch centrum sleep staging database (version 1.1),'' \url{https://doi.org/10.13026/t79q-fr32}, 2022, physioNet.

\bibitem{SPE}
\BIBentryALTinterwordspacing
C.~H. Nguyen, G.~K. Karavas, and P.~Artemiadis, ``Inferring imagined speech using eeg signals: a new approach using riemannian manifold features,'' \emph{Journal of Neural Engineering}, vol.~15, no.~1, p. 016002, dec 2017. [Online]. Available: \url{https://dx.doi.org/10.1088/1741-2552/aa8235}
\BIBentrySTDinterwordspacing

\bibitem{gan01}
\BIBentryALTinterwordspacing
A.~Creswell, T.~White, V.~Dumoulin, K.~Arulkumaran, B.~Sengupta, and A.~A. Bharath, ``Generative adversarial networks: An overview,'' \emph{IEEE Signal Processing Magazine}, vol.~35, no.~1, p. 53–65, Jan. 2018. [Online]. Available: \url{http://dx.doi.org/10.1109/MSP.2017.2765202}
\BIBentrySTDinterwordspacing

\bibitem{gan02}
\BIBentryALTinterwordspacing
I.~J. Goodfellow, J.~Pouget-Abadie, M.~Mirza, B.~Xu, D.~Warde-Farley, S.~Ozair, A.~Courville, and Y.~Bengio, ``Generative adversarial networks,'' 2014. [Online]. Available: \url{https://arxiv.org/abs/1406.2661}
\BIBentrySTDinterwordspacing

\bibitem{gan03}
Z.~Pan, W.~Yu, X.~Yi, A.~Khan, F.~Yuan, and Y.~Zheng, ``Recent progress on generative adversarial networks (gans): A survey,'' \emph{IEEE Access}, vol.~7, pp. 36\,322--36\,333, 2019.

\bibitem{gan04}
\BIBentryALTinterwordspacing
A.~G. Habashi, A.~M. Azab, S.~Eldawlatly, G.~M. Aly, and A.~Fahmy, ``Generative adversarial networks in eeg analysis: An overview,'' \emph{Journal of NeuroEngineering and Rehabilitation}, vol.~20, p.~40, 2023. [Online]. Available: \url{https://doi.org/10.1186/s12984-023-01169-w}
\BIBentrySTDinterwordspacing

\bibitem{gan05}
\BIBentryALTinterwordspacing
K.~G. Hartmann, R.~T. Schirrmeister, and T.~Ball, ``Eeg-gan: Generative adversarial networks for electroencephalograhic (eeg) brain signals,'' 2018. [Online]. Available: \url{https://arxiv.org/abs/1806.01875}
\BIBentrySTDinterwordspacing

\end{thebibliography}
\end{document}